\documentclass[prl,twocolumn,showpacs,amsmath,amssymb,floatfix]{revtex4}
\usepackage{graphicx}
\usepackage{ulem}
\usepackage{dcolumn}
\usepackage{bm}
\usepackage{amsmath}
\usepackage{latexsym}
\usepackage{amsfonts}
\usepackage{amssymb}
\usepackage{verbatim}
\usepackage[rightcaption]{sidecap}
\usepackage{color}\makeatletter

\newcommand{\Rmnum}[1]{\expandafter\@slowromancap\romannumeral  #1@}
\newcommand{\AB} {\it ab initio}

\makeatother
\begin{document}
\title{Transport of Correlated Electrons through Disordered Chains:\\
A Perspective on Entanglement, Conductance, and Disorder Averaging}
\date{\today}
\author{Daniel Karlsson and Claudio Verdozzi}
\affiliation{Mathematical Physics and ETSF, Lund University, 22100  Lund, Sweden}%
\begin{abstract}
We investigate electron transport in disordered Hubbard chains contacted to 
macroscopic leads, via the non-equilibrium Green's functions technique. We observe a 
cross-over of currents and conductances at finite bias which
depends on the relative strength of disorder and interactions. The finite-size scaling of the conductance is highly dependent on the interaction strength, 
and exponential attenuation is not always seen. 
We provide a proof that the Coherent Potential Approximation, a widely used method for treating disorder averages, fulfils particle conservation at finite bias with or without electron correlations.
Finally, our results hint that the observed trends in conductance due to interactions and disorder also appear as signatures in the single-site entanglement entropy.
\end{abstract}
%
\maketitle
In today's quest for novel electronics and
quantum-information technologies, materials with properties  largely
determined by electron correlations are
an important asset \cite{general}. Altogether,
they exhibit a wide range of nontrivial 
phenomena, making them 
excellent potential candidates to exploit for 
cutting-edge functionalities and devices.
However, materials behavior is often far from ideal because of 
uncontrolled, random inhomogeneities in the sample, 
i.e. disorder.  Disorder can greatly
affect the behavior of a solid (for example it can dramatically
alter conduction properties) and thus it should be considered in a comprehensive theoretical description \cite{ReviewDisorder}.

Significant understanding of the behavior of non-interacting electrons in disordered solids
is obtained in terms of a scaling theory of electron localization \cite{gangfour,McKinnon}. Interactions add great complexity to the picture, but the reverse is also true: describing electronic correlations in the presence of sample-to-sample statistical fluctuations
is much harder than for the homogenous case. For this, one can either resort
to straightforward but computationally 
expensive sums over configurations, or to analytical treatments of statistic fluctuations 
such as typical medium theory \cite{Dobro}  or the Coherent Potential Approximation (CPA)  \cite{CPA, Drchal}. 
Traditionally, CPA has been mostly used in static {\it ab initio} treatments
of disordered metallic alloys \cite{CPA-KKR}, but, recently, it
has also been used in non-equilibrium setups \cite{Vedayev, Kalitsov, Zhu2013}.  

On the whole, until now rigorous understanding of interacting electrons in strongly disordered 
systems has come primarily from numerical
studies \cite{Scalettar, McKinnon} in- and near-equilibrium regimes, 
by looking e.g. at linear conductances \cite{Thouless, Vojta,AriHarju}, spectral functions \cite{Vollhardt,Dobro,julia},
the degree of localization via the inverse participation ratio \cite{Wortis123}, 
or signatures in the entanglement entropy \cite{Berkovits,VFranca}. 

Out of equilibrium, the situation is less defined:
Even for ''simple'' cases such as 1D wires in a quantum transport setup
(for a recent review of work on 1D, see e.g. \cite{review1D}), many issues are only partially or not-at-all settled. 
For example,  {\it how do interactions and disorder together affect conduction in a small wire when a finite electric bias is applied? And what is their effect on the entanglement in the wire in the presence of a current? }
  
In this Letter, we use the non-equilibrium Green's functions (NEGF) technique \cite{KBE,Keldysh} to address these and related questions.
Specifically, we study electron transport through interacting disordered chains with Hubbard interactions and diagonal disorder (besides of being of fundamental interest, such systems are highly relevant for molecular electronics and quantum information).

Our main results are i) far from equilibrium,  the current
exhibits a non-monotonic behavior due to the competition of disorder and interactions; ii) for the cases considered, the interaction changes the exponential decrease of the conductance as a function of system size, typical of a non-interacting system, with a much weaker dependence;
iii) signatures of the mutual interplay of disorder and interactions can appear in the single-site entanglement entropy;
iv) CPA is particle conserving, with or without of electron correlations, and thus suitable for non-equilibrium treatments. \\
{\it Theoretical formulation.- }
We consider short, interacting disordered chains attached to two non-interacting leads. In standard notation, the Hamiltonian is
\begin{align}
 H = H_{RR} + H_{LL} + H_{CC} + H_{RC} + H_{CL},
\end{align}
where $R (L),C$ refer to the right (left) lead and central region, respectively. The lead Hamiltonian $H_{\alpha\alpha}$ ($\alpha=L,R$) is
\begin{align}
 H_{\alpha \alpha} = -J \sum _{\langle ij \rangle \in \alpha,  \sigma} c_{i\sigma} ^\dagger c_{j\sigma} + 
 \sum _\alpha b_\alpha \hat{N}_\alpha,
\end{align}
with $b_\alpha$ the (site-independent) bias  in lead $\alpha$ and
$J > 0$ the tunneling amplitude. The total number operator in lead $\alpha$ is $\hat{N}_\alpha = \sum _{i \in \alpha} \hat{n}_{i}$, and $\hat{n}_i =\hat{n}_{i\uparrow} + \hat{n}_{i\downarrow}$,
 $\hat{n}_{i\sigma} = c^\dagger_{i\sigma} c_{i\sigma}$. The chain/central region Hamiltonian
$H_{CC}$ is
\begin{align}
 H_{CC} = -J\sum _{\langle ij \rangle \in C, \sigma} c_{i\sigma} ^\dagger c_{j\sigma} + 
 \sum _{i} \epsilon _i \hat{n}_{i} + 
 U\sum _{i} \hat{n}_{i\uparrow} \hat{n}_{i\downarrow},
\label{central}
\end{align}
where the $\epsilon _i$'s are the (random) on-site energies of the chain, which account for disorder in the system. $U$ is the (site-independent) contact interaction strength in the chain. In obvious notation, the leads-chain coupling is 
\begin{align}
 H_{LC} +H_{RC}= -J (c^\dagger _{1_L} c_{1_C}+c^\dagger _{1_R} c_{L_C})+h.c.,
\label{leadHam}
\end{align}
i.e. the semi-infinite leads are connected to the ends of the chain. We take $J$ to be the same everywhere, which corresponds to transparent boundary conditions, i.e. only disorder and interactions affect the electron transmission through the chain. Also, we put $J=1$, which sets the energy scale. Furthermore, for non-interacting leads, one can solve in closed form for the NEGF in the central region via an embedding self-energy $\Sigma_{emb} = \Sigma_L + \Sigma_R$ \cite{Datta, Petri}.
To calculate steady-state properties, we evaluate the lesser $G^< (\omega)$ and the retarded $G^R (\omega)$ Green's functions in the chain: 
\begin{align}
 G^R (\omega ) &= \frac{1}{\omega + i\eta - H_0 - \Sigma ^R(\omega)} \label{GRet}\\
 G^< (\omega ) &= G^R (\omega) \Sigma^< (\omega) G^A (\omega). \label{GLesser}
\end{align}
In the most general case, the self-energy is $\Sigma = \Sigma _{HF} + \Sigma _{MB} + \Sigma_{emb} + \Sigma _{CPA}$, i.e the sum of Hartree-Fock, correlation, embedding and disorder contributions, respectively. When disorder is treated via numerical configuration averaging,
the $\Sigma _{CPA}$ is omitted. In all our calculations, the bias is applied only to the left lead, and the leads are half-filled. In the numerical configuration averaging, we study both box (uniform) disorder, $\epsilon_i \in [-W/2, W/2]$,  and binary disorder, $\epsilon_i = -W/2, W/2$. 
 For box disorder, we performed averages over at least 50 configurations, and we checked that this number is enough to produce reliable currents and densities. For binary disorder, we performed complete averages.

We use the 2nd Born Approximation (BA) to include correlation effects. The BA takes into account all diagrams of second order in the interaction and incorporates non-local effects. Exact benchmarks from 
small isolated clusters \cite{Puig,Hermann} and quantum transport setups \cite{Uimonen}
show that the BA is a versatile, overall fairly accurate approximation for low/intermediate
interaction strengths. In the BA, for local interactions, $\Sigma_{MB}$ in steady-state reads
\begin{align}
 (\Sigma _{MB})_{ij} = 
 U_{i} U _{j} G_{ij} (t)   G_{ji} (-t) G_{ij} (t), \label{2ndBorn}
\end{align}
which, by the Langreth rules, yields expressions for $\Sigma_{MB}^<$ and $\Sigma_{MB}^R$.
The density is calculated as 
$n_j =  2\int _{-\infty} ^\infty \frac{d\omega}{2\pi i} G^< _{jj} (\omega)$,
and the current through lead $\alpha$ is obtained via the Meir-Wingreen formula \cite{Meir1992}
\begin{align}
 I_\alpha \! = \! \int _{-\infty} ^\infty \! \frac{d\omega}{2\pi} \text{Tr} \! \left \{ 
 \Gamma ^\alpha (\omega) \left ( G^< (\omega) - 2\pi i f_\alpha(\omega) A(\omega) \right )  \right \}. 
 \label{Meir-Wingreen}
\end{align}
The trace is taken over the central region, $f_\alpha$ is the $(T=0)$ Fermi distribution of lead $\alpha$, and the non-equilibrium spectral function is defined as $-2\pi i A  = G^R  - G^A $, with $G^A = (G^R) ^\dagger$, and $\Gamma^\alpha = -2  \text{ Im} (\Sigma_\alpha)$. 
\begin{figure}
\begin{center}
\includegraphics[width=0.45\textwidth]{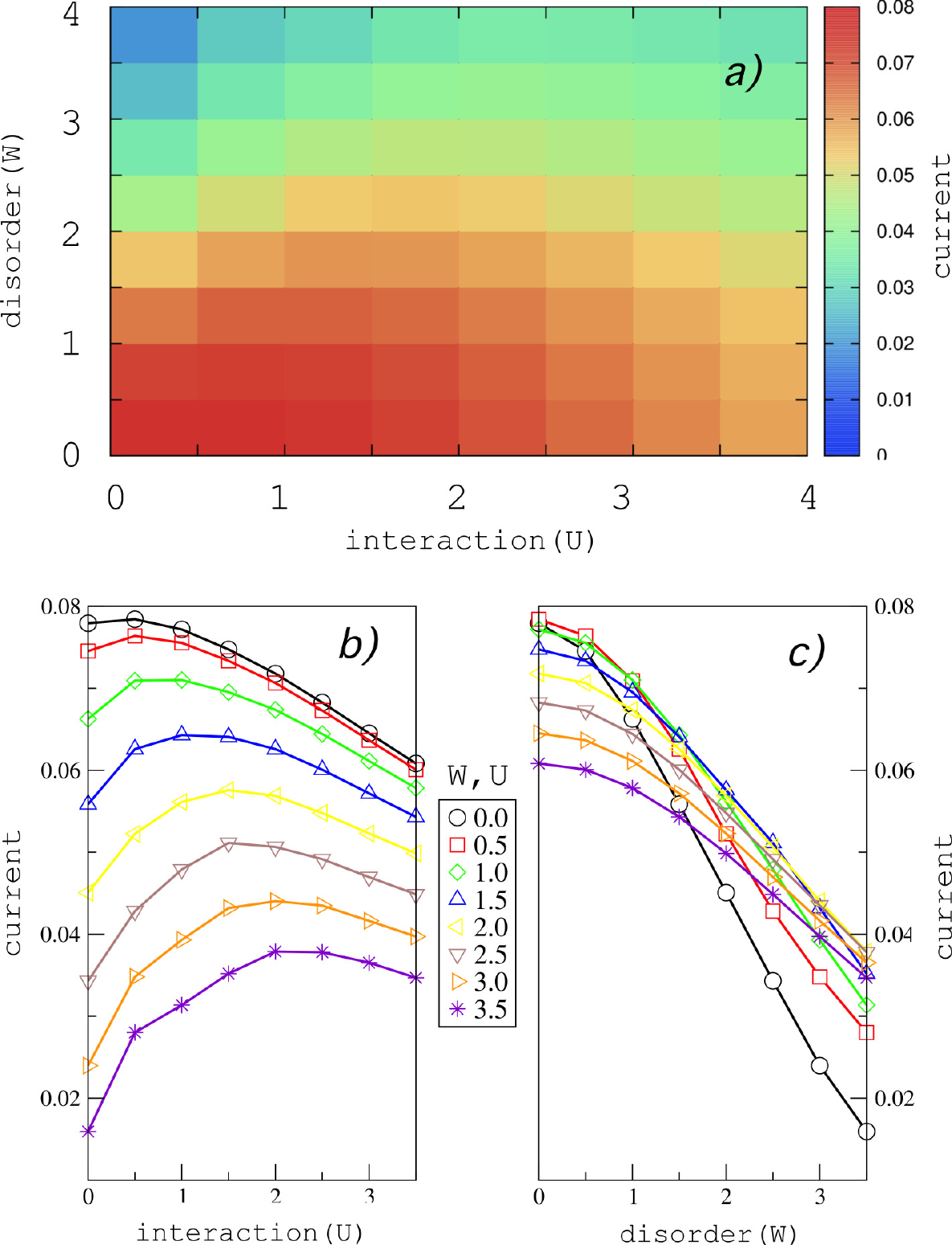}
\caption{ (Color online) The effect of interactions and disorder on the current for a bias $b_L= 0.5$. a): Heatmap for the current. b): Cuts along a fixed disorder strength (horizontal cuts in the heatmap), as a function of $U$. c): Cuts along a fixed interaction strength (vertical cuts) as a function of $W$. In b, c), the lines are a guide for the eye. The legend applies to both panels b) and c).} 
\label{conductance_heatmap}
\end{center}
\vspace{-1cm}
\end{figure}
Eqs. (\ref{GRet},\ref{GLesser}) were solved self-consistently with the 
frequency integrals in the BA performed via FFT.
\\
\noindent{\it Disorder vs interactions: results.-}
In Fig. \ref{conductance_heatmap}, we illustrate the dependence of the averaged current in the steady state on the strength of disorder $W$ and interactions $U$. Results are for a chain of $L=10$ sites, and bias $b_L=0.5$. Starting with panel a), we note that the current has a different qualitative behavior in different regions of the $U-W$ plane. In fact, close to the no-interaction (no-disorder) line the current decreases monotonically as function of disorder (interactions). However, as shown in the current heatmap, for intermediate  
interactions and/or disorder (relative the scale considered), the current clearly exhibits
non-monotonic behavior at finite bias. The behavior is clearly depicted in panel b): as a function of $W$, the region of non-monotonic behavior for the current
moves to higher $U$ values and widens. On the other hand, looking at panel c), it appears that, within the region of parameters considered, for any fixed value of $U$ the current monotonically decreases as a function of disorder strength. This observation appears to be not conclusive for higher $W$ values since, quite interestingly, the spread of the current values reduces with increasing $U$  and  it is possible to note quite distinctly that the curvature changes at high $U$ values. 

\begin{figure}
\begin{center}
\includegraphics[trim = 0.5cm 0.2cm 0.1cm 0,clip, width=0.48\textwidth]{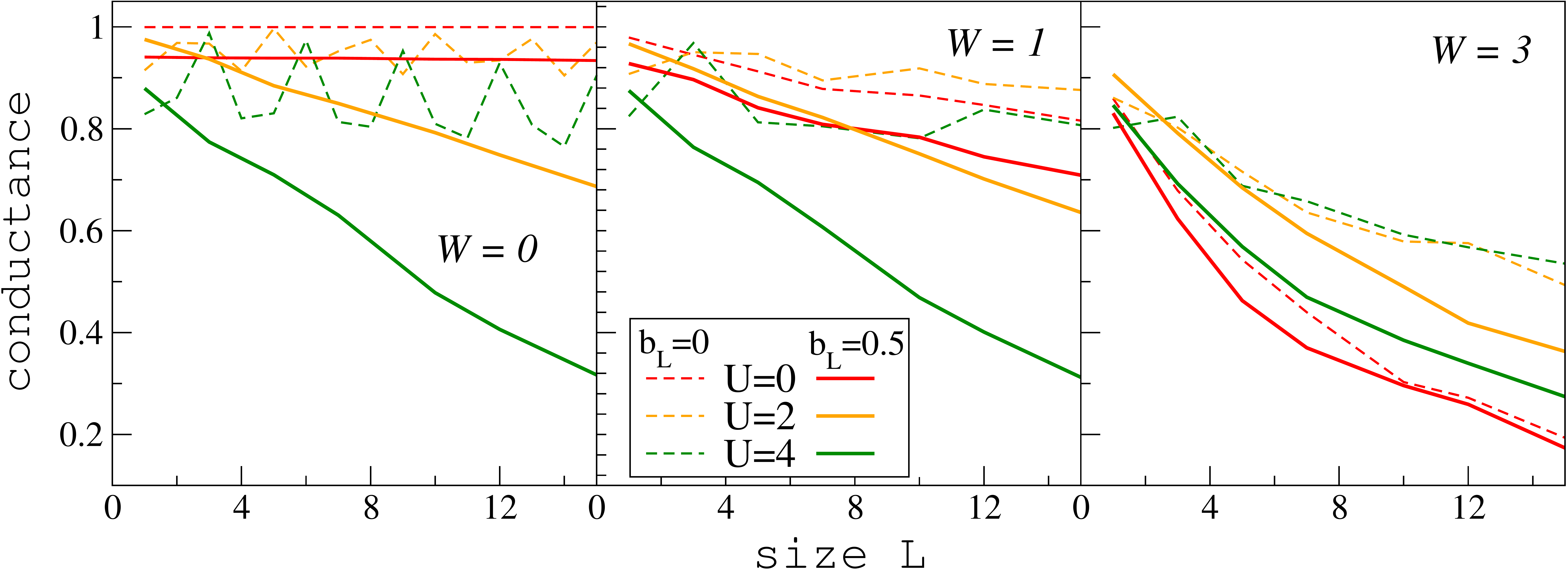}
\caption{ (Color online) Averaged differential conductance $\sigma = dI / db$ as function of the chain size $L$ for box disorder and for bias $b_L=0$ (dashed curves) and $b_L=0.5$ (solid curves). The conductance is in units of the quantum of conductance.
\label{size_scaling}}
\end{center}
\vspace{-0.6cm}
\end{figure}

The overall picture receives further support from the behavior
of the differential conductance $\sigma = \frac{\delta I}{\delta b}$, obtained by numerical differentiation. In Fig. \ref{size_scaling}, we examine how $\sigma$
scales with the size $L$ of the central region.
For $U=0$ (solid and dashed red curves), we observe the expected exponential decrease in conductance.
However, for $U>0$, $\sigma$ still decreases 
with increasing $L$ but in the interval of sizes considered, the trend is not as clear as for $U=0$. An interesting feature in the equilibrium case is that when $W=0$ and $U>0$, $\sigma$ is oscillatory. We have observed that these oscillations appear be connected to the variance of the density with the same periodicity $L=3m$ (with $m$ an integer) and, on speculative grounds, this could be related to 
Friedel oscillations induced by the lead-chain-lead boundaries when $U>0$. Coming now
to the biased case, we recover the trends discussed earlier for $L=10$, i.e.
a competition between disorder and interaction which manifests as a non-monotonic behavior
of $\sigma$ as function of $U$. This is a robust feature for $W=3$, present for all sizes considered, whilst for weaker disorder the trend in the dependence of $\sigma$ on $U$ also depends on the chain size.

Present results differ in an important way from 
a previous time-dependent DFT treatment \cite{Vettchinkina13} where non-local correlation effects were neglected: The non-monotonic trend in currents and conductances, missed in \cite{Vettchinkina13}, stems from the ability of the BA to account for such effects. 

%
\noindent {\it The Coherent Potential Approximation.-} 
We turn to another main topic of our work, namely CPA out of equilibrium. CPA treats the disorder-averaged system by an effective medium, chosen so that the average t-matrix of the local scatterer $\langle t_i(\omega) \rangle = 0$. This fulfils in an approximate way the constraint that, on average, the scattering matrix $\langle T \rangle = 0$ \cite{Soven1967}. The equilibrium CPA condition is  
\begin{align}
\langle t_i (\omega) \rangle = \left \langle \frac{ V_i - \Sigma ^{CPA}_{ii} (\omega) } {1 - (V_i-\Sigma ^{CPA} _{ii} (\omega)) G _{ii} (\omega) } \right \rangle = 0.
 \label{CPAcondition}
\end{align}
Here, $V_i$ and $ G _{ii} (\omega)$ are the impurity level and the averaged local propagator, respectively. This is how, in ground-state calculations, the complex, local in space, energy-dependent CPA self-energy $\Sigma ^{CPA}$ can be found.  

Nowadays, by combining CPA with Density Functional Theory (DFT) in a NEGF self-consistent scheme, {\AB} simulations of transport in realistic disordered systems are feasible \cite{Zhu2013, Kalitsov}. On the other hand, {\AB} NEGF treatments where CPA is combined with self-energies based on many-body approximations are still lacking. It is thus very timely and useful to assess
CPA's performance when out of equilibrium and its conserving properties as a theory. 

Usually, a conserving self-energy scheme for NEGF results 
from the existence of a so-called $\Phi$-functional \cite{Baym1962},
which guarantees particle, energy, and momenta conservation. To the best of our knowledge,
such a functional has yet to be found for the CPA, and we here take a different route
to rigorously prove that particle current is explicitly conserved out of equilibrium.
\begin{figure}
\begin{center}
\includegraphics[width=0.49\textwidth]{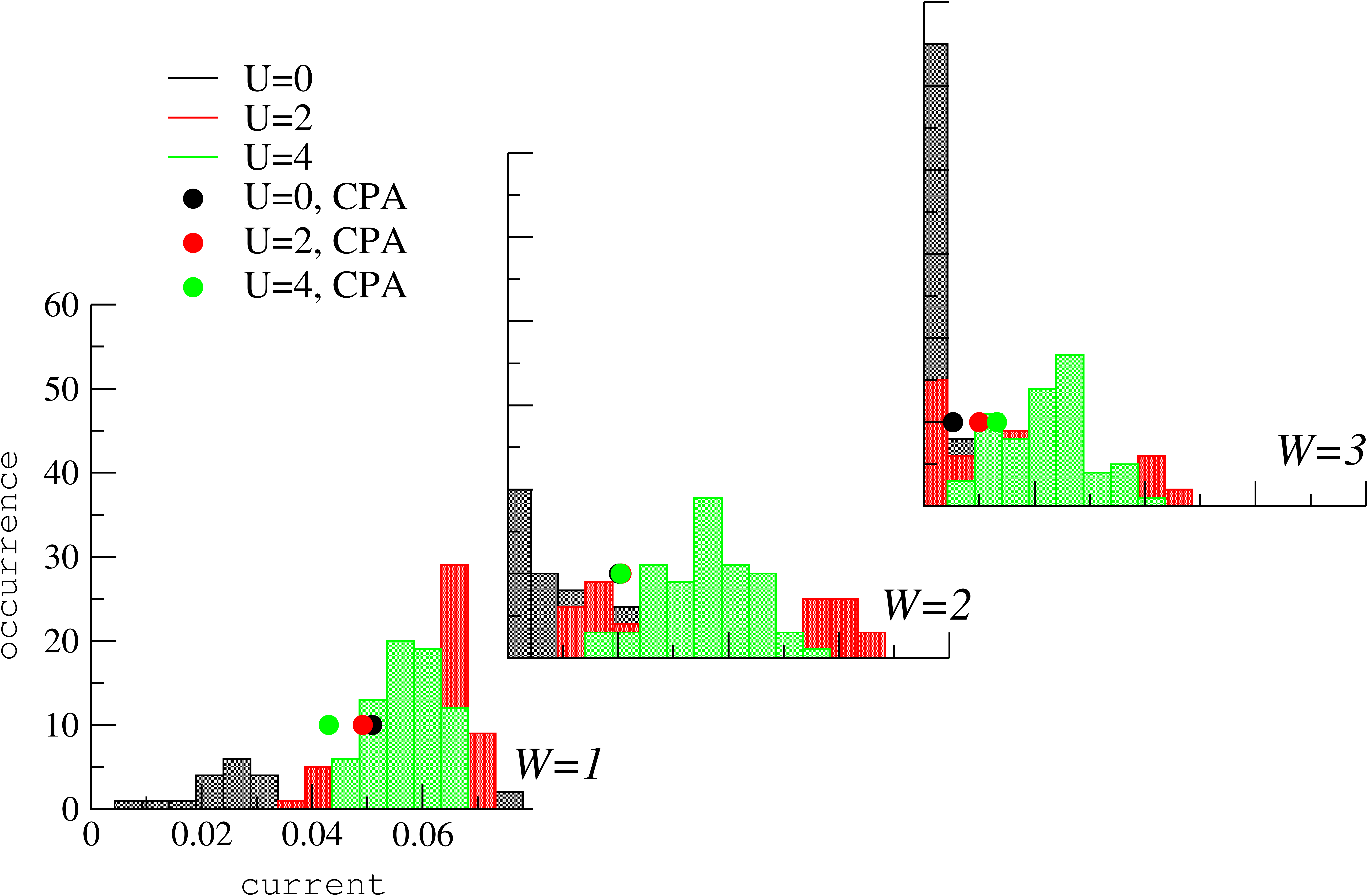}
\caption{(Color online) Steady-state currents for 8-site chains with binary disorder (A$_{50}$B$_{50}$ alloy) as a function of disorder and interaction strength.  The applied bias $b_L=0.5$.  Both results for the exact distribution of currents (histograms) and CPA current averages (circles) are shown.} 
\label{CPAcompare}
\end{center}
\vspace{-0.9cm}
\end{figure}
\begin{figure*}[t]
\includegraphics[width=0.94\textwidth]{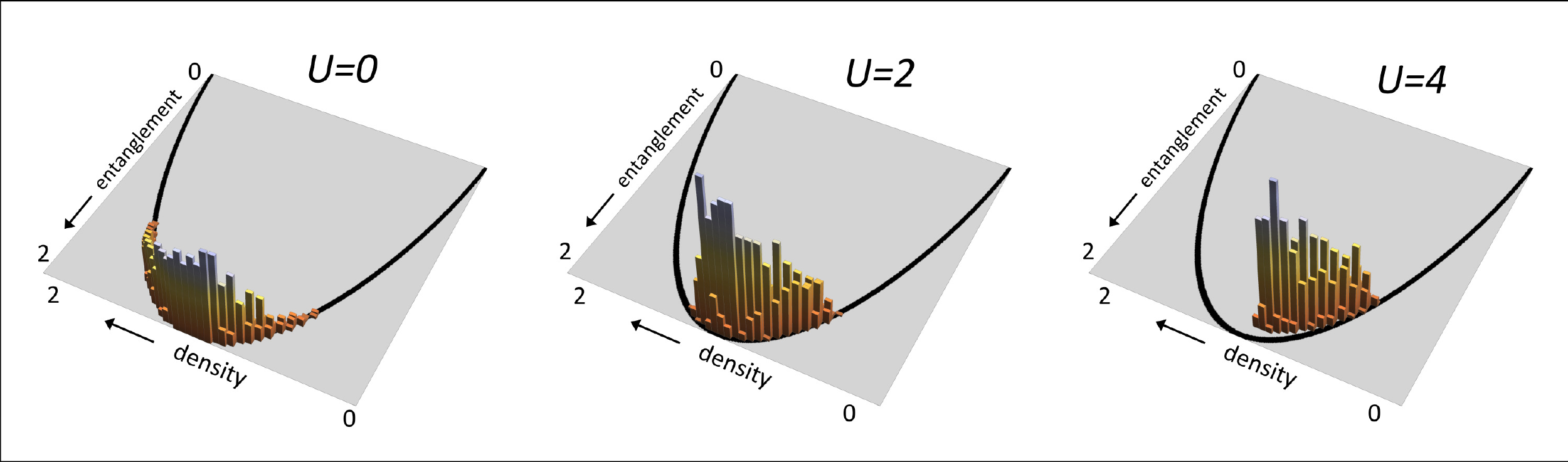}
\caption{\label{3Dplots} (Color online) 
Cumulative distribution of the single-site entanglement entropy $\mathcal{E}_k$ as a function of the interaction strength for a 10-site chain with $W=2$ and bias $b_L=0.5$. The black solid curves at the base of the entanglement histograms correspond to $\mathcal{E}_k$ for a non-interacting system, i.e. $\langle \hat{X}^\mu_{k\uparrow} \hat{X}^\nu_{k\downarrow}\rangle = \langle \hat{X}^\mu_{k\uparrow} \rangle \langle \hat{X}^\nu_{k\downarrow}\rangle$ where $n_k \in [0,2]$, and $\mathcal{E}_k \in [0,2]$.
\label{Fig3Dplot}
}
\vspace{-0.5cm}
\end{figure*}
This can be done either by reformulating Eq. (\ref{CPAcondition}) in terms of a set 
of auxiliary equations which are then reinterpreted as relations on the Keldysh contour 
\cite{Hopjan2014},
or by directly using the Langreth rules to extract the different components on the contour from Eq. (\ref{CPAcondition}).
Either way, if we schematically rewrite (site indexes and frequency arguments are 
omitted for simplicity) Eq. (\ref{CPAcondition}) as  $c =  (1-ab)^{-1} a=0$, then one can 
show that the retarded component 
$c^R = (1-a^R b^R)^{-1} a^R$ 
and,
for the lesser one, $ c^< = (1-a^R b^R)^{-1} a^< (1-a^A b^A)^{-1} + c^R b^< c^A$. 
The Langreth rules for the retarded part give the CPA condition
\begin{align}
 0 = \langle t^R \rangle  = \left \langle \frac{ V - \Sigma^R _{CPA} } {1 - (V-\Sigma^R_{CPA} )G^R } \right \rangle, \label{cpaRetarded}
\end{align}
whilst for the lesser/greater parts of $\Sigma_{CPA}$ one arrives at
\begin{align}
 \Sigma^{<,>}_{CPA} = G^{<,>} \frac{\langle |t^R|^2 \rangle} 
 {\langle | \frac{1}{1-(V-\Sigma_{CPA} ^R) G^R}|^2 \rangle}, \label{cpaLesser}
\end{align}
i.e. $\Sigma^{<,>}_{CPA} = G^{<,>}f(G^R, \Sigma^R_{CPA} )$, with $f$ a real non-negative function. 
Eq.(\ref{cpaLesser}) 
is valid for any non-correlated disorder distribution. Except for non-interacting systems,
Eq. (\ref{cpaRetarded}) and Eq. (\ref{cpaLesser}) must be solved together. 

Using Eq. (\ref{Meir-Wingreen}), the current difference $\Delta I$ becomes \cite{Thygesen08}
\begin{align}
  \Delta I = \int _{-\infty} ^\infty \frac{d\omega}{2\pi} 
  \text{Tr} \left [ \Sigma_c ^< G^>  - \Sigma_c ^>  G^<    \right ],\label{DeltaI}
\end{align}
where $\Sigma _{c} = \Sigma - \Sigma _{emb}$ refers to self-energy parts beyond the embedding self-energies. For a conserving approximation, $\Delta I = 0$. We first consider mean-field-type interactions, which means that $\Sigma ^{<,>} _c = \Sigma ^{<,>} _{CPA}$. Using Eq. (\ref{cpaLesser}), the integrand of  Eq. (\ref{DeltaI}) becomes
\begin{eqnarray}
  \sum _i (\Sigma_{CPA}^<)_{ii}  G^>_{ii}  - (\Sigma^>_{CPA})_{ii} G^<_{ii} \nonumber\\
  = \sum _i G^<_{ii} f_i G^>_{ii}  - G^>_{ii} f_i  G^<_{ii} =  0. \label{CPAconserving}
\end{eqnarray}

This holds for any number of leads, any type of (uncorrelated)  disorder, and also
when interactions are described with a mean-field type self-energy, e.g. Hartree-Fock or Kohn-Sham DFT.  To include interactions beyond mean-field, we treat the self-energies as additive,
i.e. $ \Sigma_c = \Sigma _{CPA} + \Sigma _{MB}$. The BA is, by itself, a conserving scheme.  
However, when CPA (for which the existence of a $\Phi$ is not obvious) 
and BA are combined, particle conservation needs to be proven explicitly. We again use Eq. (\ref{DeltaI},\ref{CPAconserving}),  this time only for $\Sigma _{MB}$. This is
more conveniently done in time-space (Eq. (\ref{2ndBorn})), where $\Delta I$ becomes
\begin{eqnarray}
  \Delta I \! = \! \int _{-\infty} ^\infty \! dt   
  \text{Tr} \left [ \Sigma_{MB} ^<(t) G^>(-t) - \Sigma_{MB} ^>(t)  G^<(-t) \right ]\!\!  .
\end{eqnarray}
Using the symmetries $G_{kl}^{<,>}(-t) = -[G_{lk}^{<,>}(t)]^*$, we get
\begin{eqnarray}
  \Delta I  
= 2i \sum _{kl} U_k U_l \int _{-\infty} ^\infty dt  
    \text{ Im}\left \{ [G_{kl}^<(t) ]^2   [(G_{kl}^> (t) )^*]^2 \right \}.
\end{eqnarray}
Thus $\Delta I$ is cast as a purely imaginary expression. However, all reasonable approximations give real currents and the entire expression must vanish, i.e. CPA+BA is particle conserving. 
Our numerical calculations confirm that at self-consistency $\Delta I=0$. However, 
in the initial self-consistency cycles, far away from convergence, we found that $\Delta I/I \approx 1$, i.e. self-consistency for CPA is crucial in quantum transport. \newline
Having shown the conceptual foundation of CPA for non-equilibrium treatments, 
we briefly discuss its performance in practice. By investigating 
short chains with binary disorder, we found
that CPA, at least for the systems considered, can perform rather poorly
and can in fact be unreliable even at the qualitative level.
As an example, in Fig. \ref{CPAcompare}
we report the distribution of currents for an 8-site chain with 50\% binary disorder.
As in Fig. \ref{conductance_heatmap},
increasing the interactions reduces the role of disorder,
and thus the typical value of the current (i.e. the maximum value
of the distribution) generally occurs at higher values. For $W=1,2$ it is also
true that if $U$ is further increased (i.e. $U=4$), the
current diminishes again. In any case, CPA currents are quantitatively incorrect, and only provide
the correct qualitative picture for larger disorder $W=3$.\\
\noindent {\it Entanglement, disorder, and conductance.- }
Recently, there has been an increasing interest in the use of entanglement entropy to characterize disordered interacting systems in equilibrium \cite{Berkovits,VFranca}. Here, we are interested in the non-equilibrium case, and specifically consider the single-site entanglement entropy $\mathcal{E}$, defined (in equilibrium) for site $k$ \cite{Johannensson} as $\mathcal{E}_k=-\sum_{\mu,\nu=+,-} \langle \hat{X}_{k \uparrow}^\mu \hat{X}_{k \downarrow}^{\nu}\rangle \log_2  \langle \hat{X}_{k \uparrow}^\mu \hat{X}_{k \downarrow}^{\nu}\rangle$, 
where $ \hat{X}_{k\sigma}^{+}=c^\dagger_{k\sigma}c_{k\sigma}$ and 
$ \hat{X}_{k\sigma}^{-}=1-\hat{X}_{k\sigma}^{+}$. The expression for $\mathcal{E}_k$ is straightforwardly generalized to finite biases. It is readily seen that $\mathcal{E}_k$
can be expressed in terms of the particle density $n_k
=\langle \hat{X}_{k \uparrow}^+ + \hat{X}_{k \downarrow}^+\rangle$, 
obtained via $G^<$, and the double occupancy  $ d_k=\langle \hat{X}_{k \uparrow}^+ \hat{X}_{k \downarrow}^+\rangle$. 
For the latter, we take the steady-state limit of the expression for the time-dependent
double occupancy given in \cite{PuigvonFriesen2011} and, for numerical
convenience, separate the Hartree-Fock part:
\begin{eqnarray}
 d_k  = \frac{n_k^2}{4} + \frac{1}{U_k} \int _{-\infty}^\infty \frac{d\omega}{2\pi i}
 \left ( \Sigma^< _{MB}  G^A + \Sigma^R _{MB}  G^<\right )_{kk}. 
\end{eqnarray}
We have examined different sets of $W, U,b_L$ parameters and their effect on
$\mathcal{E}_k$. A convenient way to scrutinize the behavior
is, for each set of parameters, to collect the pairs $(n_k,\mathcal{E}_k)$ from all sites and disorder configurations, and arrange them in a cumulative histogram. 
In Fig. \ref{Fig3Dplot} we show histograms for the cases $U=0,2,4$
when $W=2$ and $b_L=0.5$. In the non-interacting case, $\mathcal{E}_k$ is completely determined by $n_k$. For interacting systems, $\mathcal{E}_k (n_k)$ is multi-valued.
It is apparent that, on increasing $U$, 
the spread of densities is reduced, and the densites are shifted to 
lower values. Similarly, $\mathcal{E}_k$ 
shifts to lower values. More in general, studying other sets of parameters (not shown here), we found that the main effect of
increasing $W$ is to increase the spread of the distributions, whilst applying a bias results
in a shift of the density towards higher values. 
For each histogram in Fig. \ref{Fig3Dplot}, we also calculated the variance of $\mathcal{E}_k$,
and this is smallest for $U=2$. Since these parameters correspond to a crossover case in Figs. \ref{conductance_heatmap}-\ref{size_scaling}, 
this suggests a possible connection between the non-monotonic behavior of
currents (or conductances) and the single-site entanglement entropy, i.e. the latter
could be an indicator of the competition between disorder and interactions.

\noindent{\it Conclusions.-}
By means of NEGF, we investigated short disordered Hubbard chains contacted to leads to address questions regarding particle currents, 
conductances and entanglement in quantum transport. We find that, in the presence of an electric bias, interactions can increase the current through a disordered system connected to macroscopic leads, but increasing interactions further can decrease the current again. A finite-size scaling analysis for short chains reveals a sharp decrease of the conductance
in the pure disordered or interacting cases, but a much weaker drop away from these
limits. Our results generalize to the 
quantum transport case the qualitative equilibrium picture for uncontacted systems with homogeneous disorder and interactions, and partially support previous mean-field-type treatments for transport geometries. We have also shown  
that, out of equilibrium, the spread of entanglement entropy exhibits the same cross-over as for currents and conductances. Finally, we gave a proof that CPA out of equilibrium is particle conserving, with or without electron correlations on the level of 2nd Born. 
This puts non-equilibrium CPA on conceptually firm ground, and sets the stage for considering electron correlations and disorder on equal footing in {\AB } theories of systems out of equilibrium.

We thank Miroslav Hopjan for stimulating discussions. We also wish to acknowledge 
useful conversations with Herve Ness and Carl-Olof Almbladh.


\end{document}